# DARK MATTER, DARK ENERGY, AND ALTERNATE MODELS: A REVIEW

**Kenath Arun[*,1,2], S B Gudennavar and C Sivaram[3]**

[1]Department of Physics, Christ University, Bengaluru-560029, Karnataka, India

[2]Department of Physics, Christ Junior College, Bengaluru-560029, Karnataka, India

[3]Indian Institute of Astrophysics, Bengaluru-560034, Karnataka, India

**Abstract:** The nature of dark matter (DM) and dark energy (DE) which is supposed to constitute about 95% of the energy density of the universe is still a mystery. There is no shortage of ideas regarding the nature of both. While some candidates for DM are clearly ruled out, there is still a plethora of viable particles that fit the bill. In the context of DE, while current observations favour a cosmological constant picture, there are other competing models that are equally likely. This paper reviews the different possible candidates for DM including exotic candidates and their possible detection. This review also covers the different models for DE and the possibility of unified models for DM and DE. Keeping in mind the negative results in some of the ongoing DM detection experiments, here we also review the possible alternatives to both DM and DE (such as MOND and modifications of general relativity) and possible means of observationally distinguishing between the alternatives.

[*] Corresponding author:
e-mail: kenath.arun@cjc.christcollege.edu
Telephone: +91-80-4012 9292; Fax: +91-80- 4012 9222

# 1. Introduction

One of the most unexpected revelations about our understanding of the universe is that the universe is not dominated by the ordinary baryonic matter, but instead, by a form of non-luminous matter, called the *dark matter* (DM), and is about five times more abundant than baryonic matter (Ade et al., 2014). While DM was initially controversial, it is now a widely accepted part of standard cosmology due to observations of the anisotropies in the cosmic microwave background, galaxy cluster velocity dispersions, large-scale structure distributions, gravitational lensing studies, and X-ray measurements from galaxy clusters.

Another unresolved problem in cosmology is that the detailed measurements of the mass density of the universe revealed a value that was 30% that of the critical density. Since the universe is very nearly spatially flat, as is indicated by measurements of the cosmic microwave background, about 70% of the energy density of the universe was left unaccounted for. This mystery now appears to be connected to the observation of the non-linear accelerated expansion of the universe deduced from independent measurements of Type Ia supernovae (Riess et al., 1998; Perlmutter et al., 1999; Peebles and Ratra, 2003; Sivaram, 2009).

Generally one would expect the rate of expansion to slow down, as once the universe started expanding, the combined gravity of all its constituents should pull it back, i.e. decelerate it (like a stone thrown upwards). So the deceleration parameter $(q_0)$ was expected to be a positive value. A negative $q_0$ would imply an accelerating universe, with repulsive gravity and negative pressure. And the measurements of Type Ia supernovae have revealed just that. This accelerated expansion is attributed to the so-called *dark energy* (DE).

There are several experiments to detect postulated DM particles running for many years that have yielded no positive results so far. Only lower and lower limits for their masses are set with these experiments so far. The motto seems to be 'absence of evidence is not evidence of absence'. But if future experiments still do not give any clue about the existence of DM, one may have to consider looking forward for alternate theories (Sivaram, 1994a; 1999).

The best example of this is that of the orbit and position of Vulcan, which was theoretically inferred from the observation of Mercury orbit (Hsu and Fine, 2005). The deviation of its orbit, as predicted by Newtonian gravity, was attributed to the missing planet (DM). But the resolution of this discrepancy came through the modification of Newtonian gravity by Einstein and not by DM. This is unlike in the case of Uranus were the prediction and discovery were successful using DM (Neptune) theory (Kollerstrom, 2001).

## 2. Dark Matter
## 2.1 Observational Evidence for Dark Matter

The evidence for the existence of such non-radiating matter goes back to more than eighty years ago, when Zwicky (1937) was trying to estimate the masses of large clusters of galaxies. Surprisingly it was found that the *dynamical mass* of the cluster, deduced from the motion of the galaxies (i.e. their dispersion of velocities), in a large cluster of galaxies were at least a hundred times their luminous mass. This led Zwicky to conclude that most of the matter in such clusters is not made up of luminous objects like stars, or clusters of stars, but consists of matter which does not radiate (Zwicky, 1937).

Zwicky's observations were later confirmed by others and although he had overestimated the amount of DM it is now accepted as an established paradigm. Later observations starting about forty years ago, and continuing till now also revealed unmistakably that even individual galaxies like our Milky Way are dominated by DM (Rubin and Ford, 1970; Rubin, Ford and Thonnard, 1980). We know this for galaxies because it turns out that objects orbiting the galaxy at larger distances from the galactic centre move around more or less the same velocity as objects much closer to the centre, contrary to what is expected (Jones and Lambourne, 2004).

The rotational velocity should drop as $v_c \propto r^{-1/2}$, like in solar system. But at large distances, rotational curve becomes flat, i.e. $v_c \approx$ constant. This is valid for all spherically symmetric system and is valid at large distances. This gives information that mass is still growing even after light dies out ($M \propto R$). Indeed, as much as 90% of the galaxy mass is due to DM. This can only be accounted for if the mass progressively increases with radius as we move out further and further away from the central region. But this matter does not radiate as most of the light is from the central region. So the conclusion is that 90% of the galaxy is DM.

This seems to be universally true for all types of galaxies. Even in dwarf galaxies, the motion of their stars indicates the presence of DM (Bell and de Jong, 2001; Stierwalt et al, 2017). Even the 'missing satellite problem' (Moore et al., 1999; Nierenberg et al., 2016), which arises from numerical cosmological simulations that predict the evolution of the distribution of matter in the universe, could be attributed to the fact that many dwarfs have a huge amount of dark matter but very few stars, making them difficult to detect due to their inherent faintness.

Cosmological models predict that a halo the size of our Galaxy should have about 50 dark matter satellites with circular velocity greater than 20kms$^{-1}$ and mass greater than 300 million solar mass within a 570kpc radius. But the actual number of observed satellites is much lesser. The difference is even larger in the case of galaxy groups like the Local Group. (Klypin et al., 1999)

Recently, Beasley et al. (2016) reported measurements of ultra diffuse galaxies (UDGs) which have the sizes of giants but the luminosities of dwarfs. Deep imaging surveys of Fornax, Virgo, Coma and the Pisces-Perseus superclusters have revealed substantial populations of faint systems that were hidden from earlier surveys. Coma cluster for instance consists of galaxies with sizes similar to that of the Milky Way, but stellar luminosities similar to that of dwarfs. Measurements from a UDG (VCC 1287 in the Virgo cluster), based on its globular cluster system dynamics and size indicates a virial mass of $\sim 8 \times 10^{10}$ solar mass, yielding a dark matter to stellar mass fraction of ~3000 indicating that about 99.96% of the galaxy is dark matter (Beasley et al., 2016).

Apart from velocity distribution of galaxies and galaxy clusters, there are other evidences pointing to the presence of dark matter. Extended emission in X-ray observations of clusters of galaxies indicates presence of hot gas distributed throughout the cluster volume (Ferrari, 2008). If the gas is in virial equilibrium within the cluster we have:

$$kT \sim \frac{1}{2}m_p v^2 \qquad \ldots (2.1)$$

where the thermal velocity is ~1000kms⁻¹. This implies a temperature of $T \sim 6 \times 10^7 K$, which produces bremsstrahlung emission in X-rays. The total emission power density, integrated over all frequencies is given by:

$$\varepsilon_{ff} \quad 1.4 \times 10^{-27} Z^2 n_e n_i T^{1/2} \qquad \ldots (2.2)$$

Where, $n_e$, $n_i$ are the number density of electrons and ions respectively, $Z$ is the atomic number and $T$ is the temperature.

From X-ray observations, luminosity can be measured, which depends on density, temperature and volume of the cluster. The mass required to hold hot gas in cluster estimated requires vast amount of DM. Hot gas itself accounts for ~20% mass in rich clusters (which is several times mass of star).

Results from Chandra X-ray Observatory on the distribution of dark matter in a massive cluster of galaxies (such as Abell 2029, which consists of thousands of galaxies surrounded by a huge cloud of hot gas) indicate that the cluster is primarily held together by the gravity of the dark matter (Vikhlinin et al., 2006; Dietrich et al., 2012).

Another method to detect the presence of DM is gravitational lensing. This method provides for an alternate method of measuring the mass of the cluster without relying on observations of dynamics of the cluster (Tyson, Valdes and Wenk, 1990). All these different methods point to the currently accepted scenario that ~80% of the total amount of matter in the galaxy is a form of DM.

An important argument in favour of the existence of DM is the growth of structures. Observations suggest a bottom-up scenario, with the smallest structures collapsing first, followed by galaxies and then galaxy clusters. Cosmic Microwave Background (CMB) anisotropy measurements indicate models with predominant DM. DM hypothesis also agrees with statistical surveys of the visible structure. The gravity from DM increases the compaction force, allowing the formation of structures.

Simulations of billions of dark matter particles seem to confirm that a DM model of structure formation is consistent with the structures observed through galaxy surveys, such as the Sloan Digital Sky Survey and 2dF Galaxy Redshift Survey, as well as observations of the Lyman-alpha forest (Springel et al., 2005; Gao et al., 2005).

## 2.2 Classification of Dark Matter

An important categorization scheme for DM particles is the 'hot' vs. 'cold' classification. Hot Dark Matter (HDM) particles are those that are described by a relativistic equation of state at the time when galaxies could just start to form. And Cold Dark Matter (CDM) particles are those described by non-relativistic equation of state at the time when galaxies could just start to form. Some weakly interacting particles (like WIMPs), Supersymmetric, superstring, higher dimensions, Kaluza Klein etc. could be CDM, forming sub galactic objects first (i.e. bottom-up scenario) (Gelmini, 2006; Cheng, Chu and Tang, 2015).

This categorization has important ramifications for structure formation, and there is a chance of determining whether the dark matter is hot or cold from studies of galaxy formation. Hot dark matter cannot cluster on galaxy scales until it has cooled to non-relativistic speeds, and so gives rise to a different primordial fluctuation spectrum (Davis et al., 1985).

Warm dark matter (WDM) is another hypothesized form of dark matter that has properties intermediate between those of HDM and CDM, causing structure formation to occur bottom-up from above their free-streaming scale, and top-down below their free streaming scale. The most common WDM candidates are sterile neutrinos and gravitinos. WDM particles interact much more weakly than neutrinos. They decouple at temperatures much greater than the QCD temperature, and are not heated by the subsequent annihilation of hadronic species. Consequently their number density is roughly an order of magnitude lower, and their mass an order of magnitude higher, than HDM particles (Silk, 2000).

The cut-off in the power spectrum implied by WDM will inhibit the formation of small DM halos at high redshift. But such small halos are where the first stars form, which produce metals uniformly throughout the early universe as indicated by observations of the Lyman alpha forest.

Thus from observations the current favourite model for the universe is where the matter is mostly CDM (with a large cosmological constant, i.e. ΛCDM).

### 2.2.1 Decaying Dark Matter

Dark matter has survived until the present day, accounting for ~26% of the present energy density of the universe. It is still unknown whether these DM particles are absolutely stable or they have a finite but very long lifetime. This is a possibility since there is no theoretical basis predicting their stability.

In most models, dark matter stability is imposed ad hoc by imposing extra symmetries. Many particle physics models exist which contain unstable (very long-lived) DM particles. It is conceivable that the dark matter stability could be due to symmetry of the renormalisable part of the Lagrangian which is broken by higher dimensional operators, which could thus induce the dark matter decay (Bertone, Hooper and Silk, 2005).

Emission line-like spectral feature at energy E ~3.5 keV in the long exposure X-ray observations of a number of dark matter-dominated objects, such as the stack of 73 galaxy clusters (Bulbul et al., 2014) and in the Andromeda galaxy and the Perseus galaxy cluster (Boyarsky et al., 2014) has recently been observed. The possibility that this spectral feature may be the signal from decaying DM has sparked a lot of interest, and many dark matter models explaining this signal have been proposed.

## 2.3 What Dark Matter Cannot Be

### 2.3.1 Massive Astrophysical Compact Halo Object

The observed abundance of light elements created during the primordial nucleosynthesis can rule out the possibility that DM particles are baryonic in nature. The primordial nucleosynthesis strongly depends on the baryon-photon ratio. This is also supported by the observations of cosmic microwave background radiation.

The main baryonic candidates are the Massive Astrophysical Compact Halo Object (MACHO) class of candidates. These include brown dwarf stars, Jupiter-like planets, and 100 solar mass black holes.

Brown dwarfs are spheres of H and He with masses below 0.08 solar mass, so they never begin nuclear fusion of hydrogen. The MACHO project which analysed microlensing events from the Large Magellanic Cloud indicates that such objects can account for only about 10 – 20% of the missing DM (MACHO Collaboration, Alcock et al., 2000).

Another group, the EROS-2 collaboration does not confirm the signal claims by the MACHO group. They did not find enough microlensing effect with sensitivity higher by a factor of 2 (Tisserand et al., 2007).

These searches have ruled out the possibility that these objects make up a significant fraction of dark matter in our galaxy.

## 2.3.2 Hot Dark Matter and Neutrinos

The formation of such large scale structures raises few questions, including the very existence of large scale structures in just a few billion years, from a smooth homogeneous (uniform density) expanding universe and which objects formed first (top-down or bottom-up process).

Early analysis due to Jeans (1902) (much before discovery of Hubble), gives the balance between dissipative pressure force and attractive gravity in a medium of pressure P and density $\rho$, i.e. $PdV$  Gravitational energy. This implies that there is a minimal size $R$, for structures to grow under its own gravity, which is given by:

$$R \geq \frac{c_{sound}}{\sqrt{G\rho}} \sim \frac{(rR_{gas}T)^{1/2}}{(G\rho)^{1/2}} \qquad ...(2.3)$$

In an expanding medium, where $\rho \propto t^{-2}$, like the Friedmann-Robertson-Walker (FRW) universe, growth rate is not exponential, but follows power law. Any inhomogeneity in density, characterized as $\frac{\rho - \rho_{av}}{\rho_{av}}$  $\frac{\delta\rho}{\rho}$  $\delta$, grows under gravity with time, where $\rho_{av}$ is the average density.

The inhomogeneity described by $\delta$, are believed to have formed very early, during the inflation era. The primordial fluctuations (of scalar field) were already imprinted on it.

In expanding universe any gravitational contraction has to counteract expansion of ambient medium and pressures (also dark energy), with $|\delta| >> 1$ at end of structure formation process. In the beginning $|\delta| << 1$; when $|\delta| \sim 1$ the non linear growth imprints on CMB and this survives undistributed from recombination epoch.

So fluctuations $\Delta T/T$ (on different angular scales), must be accounted for in terms of matter-radiation interaction prior to the recombination era. Small fluctuations $(\Delta T/T < 10^{-4})$, imply the presence of dark matter. Dark matter would have decoupled much earlier, (for massive particles) and started clumping early to provide the required seeds for $\delta$. Hot DM gives early large scale structures which are contrary to observations where as cold DM implies a bottom up scenario. This therefore rules out the possibility that relativistic neutrinos could account for the missing dark matter (Springel et al., 2005; Bertone and Merritt, 2005).

The neutrino oscillations, which measure the mass difference squared, i.e. $\Delta m^2 \; m_1^2 - m_2^2$, between two species 1 and 2 (or more precisely what is obtained is the product of $\Delta m^2$ and the mixing angle, i.e. $\Delta m^2 \sin^2 2\theta$) imply that at least one of the three neutrino species has a tiny mass, possibly of the order of one or a few electron volts.

For neutrinos of given energy $E$, the oscillation length scales as, $E/\Delta m^2$. $\Delta m^2$ is typically of the order $10^{-2} - 10^{-4} \, eV^2$. Independent cosmological evidence, for instance, from the Planck 2015 temperature and polarization data suggest that the sum total of masses is less than ~0.126eV (Di Valentino et al., 2015; Hinshaw et al., 2013; Beringer et al., 2012; Sivaram and Sinha, 1974). This is also suggested by double $\beta$ decay experiments and earlier tritium decay end point analysis also implies a few electron volts (Bahcall, 1989).

Neutrinos are expected to have been produced profusely in the very initial stages of the universe, i.e. in the hot big bang. Similar to the microwave background which is the fossil remnant of the hot radiation (high energy radiation) which characterized the best dense phase of the early universe epochs (cooling with expansion) we also expect a fossil remnant of neutrinos which also now form a background with an estimated density of about $150 \, cm^{-3}$, per species, so that summed over all six species (neutrinos and anti neutrinos), we expect a fossil neutrino background with a number density of one thousand per cubic centimetre (Quigg, 2008).

Even with a neutrino number density of thousand per cc, with low mass of 0.1eV to 0.01eV implied by neutrino oscillations, would account for less than a percent of the missing DM.

### 2.3.3 Other Proposed Candidates

There are many other models for DM particles that are proposed which can be ruled out from basic astrophysical considerations. In one such alternate approach to dark matter, Drexler assumes that highly relativistic protons trapped in the halo of the galaxies due to the galaxies' magnetic field could possibly account for the yet unseen DM (Drexler, 2009). Various energetics involved in such a scenario indicate that this model is not plausible.

The energy of these highly relativistic protons required to be trapped in an orbit of radius ~30kpc in the galactic magnetic field of ~$10^{-6}$G is, $\varepsilon_P \sim 10^{16} \, eV$. To account for ~$10^{12}$ solar mass of DM in each galaxy, the number of such high energy protons required is, $n_P \sim 10^{62}$. The total number of protons in the galaxy is of the order of about $10^{67}$; i.e. one in every $10^5$ protons should be ultra-relativistic. The only source of such high energy protons is supernova explosions. The number of supernovae required to produce the required number of these protons will be $\sim 10^{15}$, which is about $10^5$ SN/year, which is much (a million times) above the observed limit.

The excess charge of $\sim 10^{62}$ relativistic protons in the halo of each galaxy required to account for the missing DM will cause a tremendous Coulomb repulsion between them (eleven orders greater than their gravitational attractions). For the gravitational attraction between Milky Way and the Andromeda galaxies to dominate, the maximum charge is constrained to be, $Q \sim \sqrt{G}M \approx 10^{51}e$, which is $\sim 10^{11}$ orders smaller than the number of relativistic protons required to account for the dark matter. Hence such a scenario is not viable (Sivaram, Arun and Nagaraja, 2011a).

Another alternate explanation for the flat rotation curve of the spiral galaxy is with an analogy with electromagnetism. Fahr (1990) has postulated a gravo-inductive force distinct from the usual Newtonian force. The exact analogy with electromagnetism assumed in postulating this force would for reason of consistency imply a vanishingly small coupling rather than the anomalously large coupling required. Hence such a gravo-inductive force cannot account for the flat rotation curve (Sivaram, 1993).

There are other earlier candidates that are ruled out by non-astrophysical considerations (Trimble, 1988). These include, majoron and goldstone boson ($10^{-5}$eV like axion), paraphoton and right-handed neutrino, keV particles from modified QCD and superweak theory, cosmion, flatino, magnito (MeV to GeV particles from SUSY and supergravity), preons, multi TeV particles, pyrgons, maximos, Planck mass particles in higher dimension theories, etc.

Other baryonic candidates that are ruled out from gravitational lensing observations; since they contribute too little to DM; include brown dwarfs, old white dwarfs, neutron stars, stellar mass black holes, solid $H_2$, dense cold molecular clouds in galaxies (which are firmly ruled out by the absence of absorption) (Clarke et al., 2004), high velocity clouds of $H_2$, and lensing also rules out stellar mass quark nuggets, boson stars, strange quark nuggets, Klein-Kaluza gravitino trapped inside neutron stars, branons, lightest Klein-Kaluza bosons, (Casse et al., 2004) etc. Studies on decaying CDM and annihilation of DM (Sigurdson and Kamionkowski, 2004; Boehm et al., 2004) also rule out many possible DM candidates.

LIMPs are DM particles that weakly interact only with leptons and have masses of 1 to 10 TeV (Baltz and Bergstrom, 2003). As the interactions are only leptonic, current elastic scattering experiments are not sensitive to this particle and these predictions rely heavily on the structure of the Galactic halo.

Another proposed DM particle is the superWIMPs (superweakly-interacting massive particles) (Feng, Rajaraman and Takayama, 2003). Such particles appear in the form of gravitinos and gravitons in theories with supersymmetry and extra dimensions. They satisfy existing constraints from the big bang nucleosynthesis and CMB but are indistinguishable in conventional DM experiments.

## 2.4 Possible Candidates of Dark Matter

There is no shortage of ideas as to what the dark matter could be. Serious candidates have been proposed with masses ranging from $10^{-5}$eV ($10^{-71}$ solar mass) to $10^4$ solar mass black holes. That's a range of masses of over 75 orders of magnitude (Zioutas et al., 2004; Redondo and Ringwald, 2011; Feng, 2010).

As we have seen, baryonic DM candidates are cosmologically insignificant. Hence much of the focus is primarily on non-baryonic candidates. The non-baryonic candidates are basically elementary particles which are either not yet discovered or have non-standard properties. There is compelling evidence that much of the DM may be made up of as yet undiscovered particles like axions, neutralinos, gravitinos or composites of the same.

### 2.4.1 Weakly Interacting Massive Particles

A fraction of a second after the Big Bang the universe was so hot that new particles (and antiparticles) were created and destroyed all the time. It turns out that a stable particle of mass near 100 GeV and interacting with the strength of the weak force will leave just about the right amount of "leftovers" to account for the observed dark matter density (Jungman, Kamionkowski and Griest, 1996).

Perhaps the most popular extension to the Standard Model, super-symmetry predicts that each particle in the Standard Model has a heavier partner of different spin but similar interactions. The lightest of these particles is stable in many cases, which is an excellent dark matter candidate (Ellis et al., 1984; McGuire and Steinhardt, 2001; Byrne, Kolda and Regan, 2002).

Many theories also suggest the possibility of more spatial dimensions that may be curled up and in which a particle could be present, which could be massive versions of the Standard

Model particles, and the lightest of these (Kaluza-Klein particle) is often stable and a good dark matter candidate (Servant and Tait, 2003).

## 2.4.2 Axions

There are other possible DM candidates which do not fit into the above framework. The most popular such candidates are called axions and arise from attempts to explain why the strong interaction seems to obey the CP symmetry (Peccei and Quinn, 1977).

Strong interactions of the standard model (QCD) possess a non-trivial vacuum structure that in principle permits violation of the combined symmetries of charge conjugation and parity, i.e. CP. Large CP violations due from the standard model would induce, an experimentally unobserved, large electric dipole moment for the neutron. This implies CP violation from QCD must be extremely tiny. The best explanation for this is the prediction of a new light neutral particle called the axion, which arises as a result of the spontaneous symmetry breaking of a new $U_a(1)$ (Peccei-Quinn) symmetry. The axion is stable, and can also be produced in the early universe (Abe, Moroi and Yamaguchi, 2002; Sivaram, 1987).

Some of the major challenges in the design of an experiment to detect axions are that the particle's mass and the coupling constant are unknown. The predicted masses range from 1μeV to 1eV. Several axion experiments are based on the prediction that axions and photons are converted into each other when subjected to a strong magnetic field.

## 2.4.3 Primordial Black Holes

Carr, Kühnel and Sandstad (2016) have considered the possibility that the dark matter could be comprised of primordial black holes (PBHs). It is found that black holes in the intermediate-mass range of one solar mass to a thousand solar mass and sub-lunar black holes in the range $10^{17} - 10^{21}$ kg can still produce all the dark matter, depending on the exact values of the astrophysical parameters involved in the constraints, including lensing, dynamical, large-scale structure and accretion. Although it is not possible to account for all of the DM in PBHs if their mass function is monochromatic but this is still possible if the mass function is extended. This perhaps requires some fine-tuning.

## 2.4.4 Exotic Candidates

In addition to the mainstream candidates above, many more exotic candidates have been suggested – WIMPzillas, gravitinos, gluinos, Q-balls, Q-nuggets, SIMPS, etc. There are myriads of possible dark matter models. One other model is that baryons can be 'packaged' in non-luminous forms. There is evidence that much of the DM may be made up of as yet undiscovered particles with several experiments all over the world trying to detect these.

Many of these conjectured particles are in the preferred range of 100 GeV to a TeV. There could be DM objects or clumps made up of these particles bound by their mutual self gravity and limits have already been placed on the abundance (density) of these objects (Sivaram and Arun, 2011a).

### 2.4.4.1 Fermi Balls

'*Fermi balls*' arise from long-range two-neutrino exchange between two electrons or protons, based on the balance between weak interactions and gravity. This force has an $r^{-5}$ dependence on the potential and has been involved in various contexts thus having a long tradition (Ivanenko and Tamm, 1934; Hartle, 1972; Sivaram, 1983).

For *N* particles in a spherical configuration of radius *R* the force is of the form (it arises purely from the Fermi-four-fermion of a neutrino pair exchanged between massive fermions with a weak-coupling strength:

$$F_W \approx \frac{G_F^2 N}{4\pi^3 R^5 \hbar c} \qquad \ldots (2.4)$$

This is to be balanced by the gravitational self energy scaling as $N^2$ with $1/R$ dependence. Balancing the four-fermion force and gravity force (overall neutrality would imply no Coulomb forces), gives a unique mass-radius (*M-R*) relation for those objects as:

$$R^4 \quad \frac{G_F^2}{\hbar c G M m} \qquad \ldots (2.5)$$

For an object of 0.1 fm radius would have mass $\sim 10^4$ kg. These objects form the Fermi balls. $10^{40}$ Fermi Balls are needed to account for the total DM in our galaxy, having a galactic number density of $10^{-20}$ m$^{-3}$. However, the radius R of these objects is much larger than their corresponding Schwarzschild radius and hence is distinct from primordial black holes (PBH).

### 2.4.4.2 Nuclear Balls

'*Nuclear balls*' could be formed in the early universe when densities were comparable to nuclear densities. Nuclear forces behave like a fluid with a surface tension of, $S_{Nucl} \approx 10^{18} Nm^{-1}$. This is a typical nuclear surface force increasing with area as $R^2$ (and with mass number as $A^{2/3}$). With the density of the nuclear fluid of $\rho_{Nucl} \approx 10^{16} kgm^{-3}$.

For the above values of *S* and ? as given by, the radii of these balls are given as: (Sivaram and Arun, 2011a)

$$R_{NB} \approx \left(\frac{S_{Nucl}}{G\rho_{Nucl}^2}\right)^{1/3} \sim 1 m \qquad \text{... (2.6)}$$

This would give them a mass $\approx 4 \times 10^{11} - 10^{12} tons$. The radius of these nuclear chunks is again much larger than their corresponding Schwarzschild radius, so they are not black holes. Of the order of $\sim 10^{27}$ of these 'nuclear balls' is required to account for the galactic DM. This implies about one such object in a volume of our solar system.

### 2.4.4.3 EW Balls and GUT Balls

At epoch earlier to the formation of 'nuclear balls', during the electroweak transition, we could have the formation of similar objects. The corresponding density would be that corresponding to the electroweak scale $\sim 10^2$ GeV, i.e. $\rho_{EW} \approx 10^{29} kgm^{-3}$ (Sivaram, 1994b; 1986a). The corresponding 'tension' is $T_{EWs} \approx 10^{29} Nm^{-1}$. The corresponding radius of this '*electroweak gravity ball*' or EW balls is:

$$R_{EWB} \approx \left(\frac{T_{EWB}}{G\rho_{EW}^2}\right)^{1/3} \sim 10^{-6} m \qquad \text{... (2.7)}$$

So these gravitating EW balls have about micron radius and weigh about 4 x $10^{11}$ kg. And as before, their radius is much larger than their Schwarzschild radius and they are thus not Hawking black holes although their mass is intriguingly close to PBH. About $10^{30}$ of these EW balls are required to account for our galactic DM which again implies one such object in a solar system volume.

We could also have such objects forming during the GUTS phase transition (Sivaram, 1990). In that case, they would be '*GUT balls*', with a tension $T_{GUT} \approx 10^{67} Nm^{-1}$ (which would depend on the GUT scale, i.e. $T \propto M_{GUT}^{-3}$) implying a much smaller radius R~$10^{-22}$ m.

### 2.4.5 New Class of Dark Matter Objects

One of the favoured dark matter candidate, WIMPs with masses from about ~ 10GeV to 1TeV, can gravitate to form a new class of objects in dark matter halos or around the galactic centre. These objects could provide the possibility of forming primordial black holes distinct from the usual Hawking black holes and they could also provide a scenario for short duration gamma ray bursts, avoiding the baryon load problem. These DM objects could also have implications for star formation (Sivaram and Arun, 2011b; Sivaram, 1994a).

These dark matter particles could form degenerate objects of planetary mass. The typical mass of such objects is given by (Sivaram and Arun, 2011b; Sivaram, 1999; Sivaram, 1994a):

$$M \approx \frac{M_{Pl}^3}{m_D^2} \quad \ldots (2.8)$$

where, $M_{Pl}$ is the Planck mass given as, $M_{Pl} \approx (\hbar c/G)^{1/2} \approx 2 \times 10^{-5} g$.

The role of DM in planetary formation and evolution has been considered by several authors. Frere, Ling and Vertongen (2008) has pointed out that local dark matter may have played a role in the formation of solar system and this raises the possibility of planet bound dark matter. It has also been recently pointed out (Diemand et al., 2008) that very concentrated dark matter clumps surviving in the solar neighbourhood could be natural nuclei for formation of stars and planets. Limits on mass density of sun-bound DM have been placed at $\sim 10^{11} (GeV/c^2) m^{-3}$ (much larger than galactic halo density) (Iorio, 2006; Sereno and Jetzer, 2006; Khriplovich, 2007).

Again Adler (2009) has suggested that accretion of planet-bound DM by Jovian planets could be a significant source of their internal heat. Indeed he points out that the low internal heat of Uranus could be because a collision (which tilted its axis) also knocked out most of its associated DM. Even earlier Sivaram (1987) had considered limits on monopole and axion fluxes from planetary heat flow measurements.

Currently there is some interest in the detection of excess gamma-rays from the galactic centre, which is attributed to the decay of 60 GeV DM particles (Huang, Zhang and Zhou, 2016). This mass for the DM particle is also favoured from other results (like DAMA experiment, among others) (Gelmini, 2006). Use of this value for $m_D$ gives:

$$M \approx 10^{26} kg \quad \ldots (2.9)$$

This is just the mass of Neptune (Sivaram, Arun and Kiren, 2016).

Trujillo and Sheppard (2014) reported the possibility of a remote massive planet (mass of Neptune) orbiting the sun; far beyond the orbit of Neptune. Batygin and Brown (2016) have inferred the presence of such a planet from the peculiar clustering of six previously known Trans-Neptunian Objects (TNOs) that have a very low (one in 15,000) probability of that happening by chance. More such objects are likely to be present, but it remains a hypothesis, until they can be detected directly. All such attempts to detect such objects (in any wavelength) have yielded negative results. So this hypothesised planet could in fact be made of DM particles.

As to the question, how many such objects could be there in the outer solar system; for the density of DM of $\sim 10^5 \, GeV/m^3$ around the solar neighbourhood, there could be one such object within about half a light year. Of course there could be many DM particles which did not form or be part of such objects.

As shown in references (Sivaram and Arun, 2011b; Sivaram, 1994a), such DM objects could have formed in the earlier epochs of the universe (when local DM density was much higher) and be in existence now. Some of them could be lurking in the outer regions of stellar-planetary systems. If such planet sized objects made of DM exists in the outer regions of our solar system, they might otherwise be undetectable optically (or in IR) and are unlikely to be present in large numbers. The above estimate shows that this may be the only such object (within a volume of the Oort cloud). So these are testable hypotheses.

It is indeed quite a coincidence that the currently favoured DM particle mass (around 60 GeV) implies a Neptune mass DM object, the number of such objects being also constrained to be one or two within a fraction of a light year.

## 2.4.6 Mirror Dark Matter

Another alternate candidate to standard DM is the mirror matter-type dark matter. It involves only a single hypothesis. Parity and time reversal symmetries stand out as the only obvious symmetries which are not respected by the interactions of the known elementary particles and these symmetries can be exact, unbroken symmetries of nature if a set of mirror particles exist. The mirror particles have the right properties to be identified with the non-baryonic dark matter in the Universe (Foot, 2004).

If mirror matter does exist in large abundances in the universe and if it interacts with ordinary matter via photon-mirror photon mixing, then this could be detected in dark matter direct detection experiments such as DAMA/NaI and its successor DAMA/LIBRA.

Mirror dark matter halos around spiral galaxies lose energy via dark photon emission. Astrophysical considerations give a lower limit on the kinetic mixing strength, and indeed lower limits on both nuclear and electron recoil rates in direct detection experiments can be estimated, and all of the viable parameter space will be probed in forthcoming XENON experiments including LUX and XENON1T. Thus these experiments will provide a definitive test of the mirror dark matter hypothesis (Clarke and Foot, 2017).

## 2.5 Detection of Dark Matter

Several experiments running for many a year have yielded no positive results so far. Only lower and lower limits for their masses are set with these experiments so far. Particles with right property could be discovered in LHC. If Weakly Interacting Massive Particles account for the DM in our galaxy, it would imply that billions of these particles (WIMPs) must pass through every square centimetre of the Earth each second.

At the Large Hadron Collider (LHC), the energy is of the order of TeV, mimicking the conditions of the universe when it was a picosecond old. The LHC has the possibility of producing and detecting DM particles that could have been produced during the early universe. These particles are expected to be pair produced in association with a Standard Model particle. The presence of the WIMPs pair is inferred from the Missing Transverse Energy which is the vector sum of the imbalance in the transverse momentum plane recoils a Standard Model particle. The collider is able to produce light mass DM which the traditional detection fails to detect due to the small momentum transfer involved in the interaction (Hoh, Komaragiri and Wan Abdullah, 2016).

The CMS (CMS Collaboration, Chatrchyan et al., 2008) and ATLAS (ATLAS Collaboration, Aad et al., 2008) experiments at the LHC have searched for new particles in proton-proton collisions at a centre-of-mass energy of 7 TeV. CMS and ATLAS have studied a number of new particle signatures by scanning the parameter space of different supersymmetric and extra-dimensions models. The presence of a dark matter particle would only be inferred by observing events with missing transferred momentum and energy.

The usual method of detecting DM particles is by their collision with nuclei (of various detectors like CdTe, Xe, etc.) and the transfer of energy to the nucleons (which thereby recoil and this can be measured) (Undagoitia and Rauch, 2016).

Another method is using Superconducting Super Granules (SSG) (Pretzl, 1993). For example a small 10micron diameter sphere of tin kept at just the superconducting transition temperature, $T_C$ (say kept at 500mK) if hit by a 30eV particle, temperature can rise by 20mK, enough to destroy the superconducting state (in tin granule).

So this is a sensitive method. Similarly vortices in superfluid He can also dissipate if energy from a DM particle is absorbed. Many of the ongoing detectors are cryogenic (Agnese et al., 2016).

These experiments can be divided into two classes. The direct detection experiments searches for the scattering of DM particles off atomic nuclei, while the indirect detection experiments looks for the products of annihilations of the DM particles.

**2.5.1 Direct Detection Experiments**

Direct detection experiments usually operate in deep underground laboratories to reduce the background from cosmic rays. These include: the Soudan mine; the SNOLAB underground laboratory at Sudbury (Canada); the Gran Sasso National Laboratory (Italy); the Canfranc Underground Laboratory (Spain); the Boulby Underground Laboratory (UK); the Deep Underground Science and Engineering Laboratory (US); and the Particle and Astrophysical Xenon Detector (China).

The majority of present experiments use either a cryogenic detectors. They detect the flash of scintillation light produced by a particle collision in liquid xenon or argon (Lee et al., 2014).

The Large Underground Xenon experiment (LUX) is another experiment that aims to directly detect WIMP dark matter particles through their interactions with ordinary matter. The detector contains 370 kg liquid xenon detection mass which identify individual particle interactions (Beringer et al., 2012; LUX Collaboration, Akerib et al., 2016).

LUX is designed to search for WIMPs of around ten to hundred times the proton mass, which interact very weakly (interaction cross-section being trillions of times smaller than that for photons) with the Xenon nuclei in the detector. If one of the WIMPs collides into one of these densely packed Xenon atoms in the detector, it would produce electroluminescence photons.

After examining a huge amount of data from carefully calibrated devices over a twenty month period the latest results reveal that nothing with the right properties to excite the Xenon nuclei made it through, i.e., a negative result. A less sensitive experiment ending in 2013 also ended with a negative result. This upgraded LUX detector has given the best sensitivity signal so far since the earlier 2013 result (LUX Collaboration, Akerib et al., 2017).

After this recent negative result of the LUX two year run, the focus is now on the large Xenon1T experiment currently started in Italy at the underground Gran Sasso laboratory. It is now the largest most sensitive search for WIMPs, and uses ten times as much xenon as the LUX experiment. It houses a cylindrical vessel of 3500 kilogram (three and a half tons) of liquid Xenon located 1400 metres underground (XENON Collaboration, Aprile et al., 2015).

In principle it should detect 300 events per year, with its large mass of liquid Xenon detector. The experiment is planned for two years and detection of even ten particles that match (or appear to match) the predicted property of the expected DM is considered enough to claim a discovery.

**2.5.2 Indirect Detection Experiments**

Indirect detection experiments search for the products of WIMP annihilation or decay. A WIMPs could annihilate with another (if these particles are their own antiparticle) to produce gamma rays or other particle-antiparticle pairs that can be detected. Additionally, if these particles are unstable, they could decay into standard model particles (Choi, Kyae and Shin, 2014). These processes could be detected indirectly through an excess of gamma rays, antiprotons or positrons emanating from regions of high dark matter density.

The Fermi Gamma-ray Space Telescope, launched 11 June 2008, is searching for gamma rays from dark matter annihilation and decay (Carson, 2007). A few of the WIMPs passing through the Sun or Earth may scatter off atoms and lose energy. This way a large population of WIMPs may accumulate at the centre of these bodies, increasing the chance that two will collide and annihilate. This could produce a distinctive signal in the form of high-energy neutrinos originating from the centre of the Sun or Earth.

The Fermi-Large Angle Telescope (Fermi-LAT) probes photons of the highest energies. At such energy scales, these particles may exhibit signatures of the new physics, which deviate significantly from the Standard Model. In some of these models, the dark matter particle may self-annihilate or decay into standard model particles, including photons with energies as large as the dark matter particle rest mass. The detection of secondary gamma-rays from these processes with the Fermi-Large Angle Telescope (LAT) could provide compelling evidence for the dark matter particle.

Dwarf spheroidal galaxies can represent a very clean system to search for dark matter annihilation. Currently, there are roughly 25 known dwarf satellite galaxies to the Milky Way, and both ground-based instruments such as high-energy stereoscopic system (H.E.S.S.), major atmospheric gamma-ray imaging Cherenkov (MAGIC), and very energetic radiation imaging telescope array system (VERITAS), the Fermi-LAT, and the future CTA are actively observing these objects (Funk, 2015).

It is generally considered that the detection of such a signal would be the strongest indirect proof of WIMP dark matter. High-energy neutrino telescopes such as AMANDA (Antarctic Muon And Neutrino Detector Array), IceCube (Abbasi et al., 2009) and ANTARES (Astronomy with a Neutrino Telescope and Abyss environmental RESearch project) (Adrián-Martínez et al., 2014) are searching for this signal. Gamma rays from galaxy centre (due to WIMP–anti-WIMP annihilation) are also being looked for by using Payload for Antimatter Matter Exploration and Light-nuclei Astrophysics (PAMELA).

For axion detection special methods are invoked, like a microwave cavity containing a magnetic field. Axions would have been produced in the early Universe by the vacuum realignment mechanism and radiation from cosmic strings, leading to a cold dark matter component, as well as from thermal interactions, leading to a hot dark matter component.

Axions would also emerge from the hot interiors of stars, the Sun being the most powerful 'local' source (Raffelt, 2008). To search for these axions, one can use magnetically induced $a\gamma$ conversion in a dipole magnet pointing toward the Sun ('axion helioscope' technique (Sikivie, 1983)).

Solar axions are expected to be in X-ray range, through the conversion $a \to 2\gamma$. If DM is dominated by axions (in the galactic halo for instance) their mass will be of the order of $10^{-5}$ eV (Beck, 2013).

The Axion Dark Matter eXperiment (ADMX) uses a resonant cylindrical microwave cavity within a large superconducting magnet. If the axions turn out to have low masses, they may show up as microwaves detectable by the microwave cavity. The experiment can detect low-mass axions in the range of 1 μeV to 10 μeV (Asztalos et al., 2010).

PVLAS ("polarization of the vacuum with laser") aims to test of QED and as a result detect these DM particles by using a high-finesse Fabry-Perot optical cavity (Zavattini et al., 2012).

The CERN Axion Solar Telescope (CAST) is searching for axions originating from the sun. If axions exist, they may be produced in the Sun's core when X-rays scatter off electrons and protons in the presence of strong electric fields. So far, CAST has narrowed down the range where these particles may exist and has set limits on axion coupling to electrons and photons, but it has not turned up definitive evidence for solar axions (Barth et al. 2013; Arik et al. 2011).

## 3. Dark Energy

Various observations of the dynamics of the universe have implied the dominance of DE. This has led to the introduction of a repulsive gravity source to make the deceleration parameter negative (Jones and Lambourne, 2004). The dimensionless quantity, deceleration parameter $q$ measures the cosmic acceleration of the universe's expansion:

$$q = -\frac{\ddot{a}a}{\dot{a}^2} \qquad \text{... (3.1)}$$

where '$a$' is the scale factor of the universe.

All postulated forms of matter yield a deceleration parameter $q \geq 0$ (positive $q$), except in the case of DE. Therefore, any expanding universe is expected to have a non-increasing Hubble parameter. The observations of the CMB demonstrate that the universe is close to flat, i.e.:

$$q = \frac{1}{2}(1+3w) \qquad \text{... (3.2)}$$

For any cosmic fluid with equation of state $w$ greater than − 1/3, the universe is decelerating. Where, the equation of state parameter is defined by the ratio of the pressure to the corresponding energy density, i.e. $w = P/\rho$.

However, observations of distant supernovae (Perlmutter et al., 1999; Riess et al., 1998) indicate that $q$ is negative, i.e. an accelerated expansion of the universe. This indicates the existence of dark energy that dominates at the current epoch. DE (negative pressure), causes repulsive gravity, and hence restricts the size of structures, which indicates that it was less dominant during structure formation. The models for DE range from a cosmological constant (lambda) term to quintessence, Chaplygin gas, etc. (Sivaram and Sinha, 1979).

### 3.1 Cosmological Constant

The cosmological constant is the energy density of vacuum, originally introduced by Einstein (1917) as an addition to his theory of general relativity to make the universe static. Einstein abandoned the concept after Hubble's discovery that all galaxies outside the Local

Group are moving away from each other. Pressure due to a cosmological constant term, $\Lambda$ is given by (which is valid for any curvature):

$$P_{-ve} \approx -\frac{\Lambda c^4}{8\pi G} \qquad \text{... (3.3)}$$

If just $\Lambda$ remains constant, matter density scales as $(1+Z)^3$. So at $Z \sim 1$, matter density becomes equal to the DE density. The pressure is given by, $P = w\rho c^2$, $w = -1$ gives cosmological constant $\Lambda$.

The dark energy density is given by, $\rho \propto \rho_0 a^{-3(1+W)}$, where, $\rho_0$, the initial value may be related to $\Lambda$ as $\frac{\Lambda c^4}{8\pi G}$. The general solution for a dark energy dominated universe is:

$$a(t) = \left[ a_0^{3(1+W)/2} + \frac{3}{2}(1+W)\left(\frac{8\pi G \rho_0}{3}\right)^{1/2} t \right]^{2/[3(1+W)]} \qquad \text{... (3.4)}$$

For $w = 0$, we recover the matter dominated universe, $\rho = \rho_0 a^{-3}$, and for $w = 1/3$, we have the radiation dominated universe: $\rho = \rho_0 a^{-4}$

### 3.2 Quintessence

Quintessence is a model for dark energy, proposed by Peebles and Ratra (1988), as an alternative to the cosmological constant.

It is proposed to be a fifth fundamental force. The non-zero value of $\Omega_\Lambda = 0.7$ has many problems, such as:

1. The value of vacuum density, $\rho_{vac} = (10^{-2} eV)^4$, is unnaturally small. Even the electroweak or super-symmetry breaking at TeV scale gives $\rho_{vac} > (1 TeV)^4$, the observed value is $10^{54}$ times smaller.

2. At present $\Omega_m$ (scaling as $R^{-3}$) and $\Omega_\Lambda$ (scaling as $R^0$) are of same order of magnitude, implying we live in a very special era.

This model tries to address these problems (Wetterich, 2014). In this model, an inflation field is so tailored that, $T_{\mu\nu}(t)\langle\phi\rangle = \Lambda(t)g_{\mu\nu}$, $\Lambda(t)$ is a function of time and the potential $V(\phi)$ can be $V(\phi) \sim M^{4+\beta}\phi^{-\beta}$ or $V(\phi) = M^4[\exp(M/\phi)-1]$, where $M$ is a parameter. By arranging that $\rho_\phi$ is a little below $\rho_\gamma$ at the end of inflation, it can track $\rho_\gamma$ and then $\rho_m$ (after recombination), such that $\Omega_{\Lambda(t_0)} \sim \Omega_m$. The action is given by:

$$S \quad \int d^4x \sqrt{-g} \left[ \frac{1}{2} \partial_\mu \phi \partial^\mu \phi - V(\phi) \right] \qquad \text{... (3.5)}$$

The quintessence pressure and corresponding energy density are:

$$\rho_Q \quad \frac{1}{2}\dot\phi^2 + V(\phi)$$
$$P_Q \quad \frac{1}{2}\dot\phi^2 - V(\phi) \qquad \text{... (3.6)}$$

And the equation of state parameter is:

$$w_Q \quad \frac{\dot\phi^2 - V(\phi)}{\dot\phi^2 + V(\phi)} \qquad \text{... (3.7)}$$

A special case of quintessence is the k-essence, which has a non-standard form of kinetic energy (Rozas-Fernández, 2012). The kinetic variable is given by:

$$X \quad \frac{1}{2} \partial_\mu \phi \partial^\mu \phi \qquad \text{... (3.8)}$$

And the Lagrangian general function of the kinetic energy term is:

$$L_K \quad \int d^4x \sqrt{-g} P(\phi, X) \qquad \text{... (3.9)}$$

These models claimed not to require fine tuning (Zlatev, Wang and Steinhardt, 1999). This is controversial since Kolda and Lyth (1999) claim that slow-roll inflation and quintessence require fine tuning to a level of 1 part in $10^{50}$.

### 3.3 Phantom Energy

Phantom energy is another hypothesised form of DE, having a negative kinetic energy that increases with the expansion of the universe. Due to which it could cause the expansion of the universe to accelerate so quickly that it will lead to a 'Big Rip' (Caldwell, Kamionkowski and Weinberg, 2003).

It is also possible that the equation of state parameter $w < -1$. Data (from Planck data) from CMB and supernova limit the range to $-1.19 < w < -0.95$ (Ade et al., 2014; Kumar and Xu, 2014).

A $w < -1$, implies that DE density is growing with time. This has drastic implications for future evolution of the universe and stability of gravitationally bound structures. Increase in dark energy density will ultimately strip apart gravitationally bound objects and rip or disrupt all bound structures, even ultimately atoms, nuclei, etc. (Big Rip).

In a cosmological constant Universe, the scale factor grows more rapidly than the Hubble distance and galaxies will begin to disappear beyond the horizon. With phantom energy, the expansion rate grows with time, the Hubble distance decreases, and the disappearance of galaxies is accelerated. The current data indicate that our Universe is poised near the separation between phantom energy, cosmological constant, and quintessence.

## 3.4 Quintom Dark Energy

The Quintom DE (quintessence and phantom) is another hypothetical scenario regarding dark energy, with a time varying equation of state parameter $w(z)$, that can cross the phantom divide of $w \approx -1$. This parameter is given by:

$$w_{DE} \approx w_0 + w'(z) \qquad \ldots (3.10)$$

which is valid at low z; and at high z it is given by:

$$w_{DE} \approx w_0 + w_1(1-a) \approx w_0 + w_1 \frac{z}{1+z} \qquad \ldots (3.11)$$

where, $a$ is the scale factor and $w_1 \approx -dw/da$

The measurement of the parameter (Serra and Dominguez Romero, 2011) indicates:

$$w_0 \approx -1.06 \pm 0.14, \quad w_1 \approx 0.36 \pm 0.62 \qquad \ldots (3.12)$$

Current observational data mildly favours $w_{DE}$ crossing the phantom divide during evolution. But ΛCDM is still in great agreement with observations (Upadhye, Ishak and Steinhardt, 2005; Marciu, 2016).

The problem is that it cannot account for why the Universe entered the cosmic super-acceleration at the present epoch. Theoretically this model is a no-go theorem. Single fluid or a scalar field cannot realise a viable Quintom model in FRW. Either a non-minimal coupling has to be introduced or extra degrees of freedom. For detailed discussion of the model see for example Cai et al. (2010) and references there-in.

## 3.5 Dark Energy and Mach's Principle

Mach's Principle implies that the local standards of non-acceleration are determined by some average of the motions of all the masses in the universe. As a result, it even implies interactions between inertia and electromagnetism (Rindler, 1977).

In the case of gravity similar force has been proposed (Einstein-Sciama force) for mass $m_1$ and $m_2$ separated by $r$:

$$F_{Grav} \quad \frac{Gm_1m_2}{r^2} + \frac{Gm_1m_2}{rc^2}a \quad \ldots (3.13)$$

And for large r, the $1/r$ term dominates, i.e.

$$F_{Grav} \approx \frac{Gm_1m_2}{rc^2}a \quad \ldots (3.14)$$

If at all gravitational forces contribute to local inertia, this term dominates. Thus for the cumulative effect of all distant masses $m_i$ on the local mass $m$,

$$F_{Grav} \quad \sum_i \frac{Gm_i}{rc^2}ma \quad \ldots (3.15)$$

which is summed over all the masses $m_i$

$$\sum m_i \quad \int_{Vol} \rho dV \quad \ldots (3.16)$$

$\rho$ is the average density of matter in the universe. From this we have:

$$F_{Grav} \quad \frac{G}{c^2}\left[\int_{Vol}\frac{\rho dV}{r}\right]ma \quad \frac{G}{c^2}\left[\int_0^{R_H}\frac{\rho}{r}(4\pi r^2 dr)\right]ma \quad \ldots (3.17)$$

(The integration carried out over the Hubble volume)
This gives:

$$F_{Grav} \quad \frac{2\pi G\rho R_H^2}{c^2}ma \quad \frac{2\pi G\rho}{H_0^2}ma \quad \ldots (3.18)$$

Where

$$H_0^2 \quad \left(\frac{c}{R_H}\right)^2 \quad \frac{8\pi G\rho_C}{3} \quad \ldots (3.19)$$

$\rho_C$ is the critical density.

Therefore we have, $F_{Grav} \quad 0.75\,ma$

If the universe is vacuum dominated or dark energy dominated, the simplest case being that of a cosmological constant, $\Lambda$ term, for which we have (Spergel et al., 2003):

$$\rho \quad \frac{\Lambda c^2}{8\pi G}; H_0^2 \quad \frac{\Lambda c^2}{3} \quad \ldots (3.20)$$

This gives:

$$F \simeq \frac{2\pi G \left(\frac{\Lambda c^2}{8\pi G}\right)}{\frac{\Lambda c^2}{3}} ma \simeq 0.75 ma \qquad \text{... (3.21)}$$

This indicates that gravitating vacuum energy (DE) could contribute up to 75% of the inertial mass of particles. The quantum vacuum energy density for the curved space (of constant curvature) is given by (Sivaram, 1986b; 1986c):

$$\rho_{Vac} \simeq \frac{\hbar c}{4\pi} \Lambda \int_0^{k_{max}} k\, dk + \hbar c \Lambda^2 \int_0^{k_{max}} \frac{dk}{k} + \hbar c \Lambda^3 \int_0^{k_{max}} \frac{dk}{k^3} + \ldots \qquad \text{... (3.22)}$$

Since $\Lambda \approx 10^{-56} cm^{-2}$, the terms with higher powers of $\Lambda$ are much smaller. They become significant only at Planck scale. Therefore we have (Sivaram, 1986a):

$$\rho_{Vac} \simeq \frac{\hbar c}{4\pi} \Lambda \int_0^{k_{max}} k\, dk \qquad \text{... (3.23)}$$

If $k_{max} \simeq \left(\frac{c^3}{\hbar G}\right)^{1/2} \sim \frac{1}{L_{Pl}}$ is the Planck cut off wavenumber, we have:

$$\rho_{Vac} \simeq \frac{\hbar c}{8\pi} \Lambda \left(\frac{c^3}{\hbar G}\right) \simeq \frac{\Lambda c^4}{8\pi G} \qquad \text{... (3.24)}$$

Using this we have:

$$F \simeq \frac{G}{c^2}\left[\frac{\hbar c}{4\pi}\Lambda \int_0^{k_{max}} k\, dk \Big/ c^2\right]\left[\int_0^{R_H} \frac{4\pi r^2 dr}{r}\right] ma \qquad \text{... (3.25)}$$

Which gives: $F \simeq \left(\Lambda R_H^2\right) ma$

This implies that $\Lambda R_H^2 \sim 1$, in order that all the local inertia is accounted for by the total gravitational interaction energy of the background vacuum energy up to Hubble radius, and hence, we have:

$$\Lambda \simeq \frac{1}{R_H^2} \approx 10^{-56} cm^{-2} \qquad \text{... (3.26)}$$

At a particular redshift, say $z \simeq z_0$, both matter and DE would have been equally significant.

$$\rho_{\Lambda_0}(1+z_0)^2 \quad \rho_{m_0}(1+z_0)^3 \Rightarrow (1+z_0) \approx 3 \qquad \ldots (3.27)$$

That is at a redshift of ~2, which is consistent with observations as it has been observed that $\Lambda$ is dominant even at about 8 billion years ago corresponding to $z_0 \approx 2$ (Hicken et al., 2009). At primordial nucleosynthesis, that is at $z \approx 10^9$, the energy densities are given by (Kolb and Turner, 1990):

$$\rho_\Lambda \approx 10^{18} \rho_{\Lambda_0}, \rho_m \approx 10^{27} \rho_{m_0}, \rho_R \approx 10^{36} \rho_{R_0} \qquad \ldots (3.28)$$

Thus at nucleosynthesis, the dark energy density was several orders lower than radiation energy density. So it will not affect element abundance. At recombination, that is $z \approx 10^3$ the energy densities are given by:

$$\rho_\Lambda \approx 10^6 \rho_{\Lambda_0}, \rho_m \approx 10^9 \rho_{m_0}, \rho_R \approx 10^{12} \rho_{R_0} \qquad \ldots (3.29)$$

So $\rho_m$ and $\rho_R$ at recombination are several orders higher than $\rho_\Lambda$. So it will not much affect CMBR. Only at around $z \approx 2$, $\rho_\Lambda$ would dominate (Sivaram and de Sabbata, 1993; Sivaram and Arun, 2013).

Sivaram and Sinha (1975; 1976) were perhaps the first to consider a time varying cosmological constant with $\Lambda$ varying as $1/R_H^2$ with epoch. Also, a time dependent vacuum energy with, $\Lambda \propto t^{-2}$ was obtained from a theory of unification of gravity with other interactions. It was shown that the quantum vacuum energy owing to spontaneous symmetry breaking of a generalised gauge theory, in the early universe, varies with epoch as $t^{-2}$. An exact solution was given as (Sinha, Sivaram and Sudershan, 1976):

$$R \quad R_0 \cosh^{3/2} \sqrt{\Lambda} t \qquad \ldots (3.30)$$

This paper also gives an inflationary exponential expansion driven by a large cosmological vacuum energy in the early universe.

$$\Lambda_{Present} \approx \Lambda_{Pl}\left(\frac{t_{Pl}}{t_H}\right)^2 \approx 10^{-56} cm^{-2} \qquad \ldots (3.31)$$

This is just the value deduced for the dominant dark energy density at present (Sivaram, 2009).

### 3.6 Gurzadyan – Xue Dark Energy

Gurzadyan and Xue (2003) suggests the possibility of variation of physical constants such as the speed of light and the gravitational constant, which provides a value for DE density in remarkable agreement with current cosmological observations, unlike numerous phenomenological scenarios where the corresponding value is postulated. Many recent papers (Cuesta et al., 2008) give the Gurzadyan-Xue DE formula as:

$$\rho_{GX} \approx \frac{\pi}{8} \frac{c^4}{G} \frac{1}{a^2} \qquad \text{... (3.32)}$$

Sivaram (1986a; 1986c) has derived the identical formula much earlier, which has a detailed derivation based on the higher curvature power expansion for the vacuum fluctuation energy. The zero point energy associated with the curving of background space can be expressed as the action by expanding the Lagrangian as a series in powers of the curvature. The vacuum energy density is given by:

$$\rho_{vac} \approx \frac{1}{2} \hbar c R N^2 k_0^2 \qquad \text{... (3.33)}$$

Where, $R \approx R_H^{-2}$, $N^2 k_0^2$ is the Plank wave number $\frac{c^3}{\hbar G}$; so the above equation gives:

$$\rho_{vac} \approx \frac{1}{2} \frac{c^4}{G} \frac{1}{R_H^2} \qquad \text{... (3.34)}$$

The dark energy $\rho_{vac}$ is then given as $10^{-8} ergs/cm^3$ or $10^{-29} gm/cm^3$, exactly what is now implied by the present observations.

## 4. Dark Matter and Dark Energy

As of today, we don't know if dark matter and dark energy are manifestations of the same dark "thing". For now, we think of them as separate entities. But the difference between the two is in the pressure exerted by them. The dark energy and cosmic repulsion is associated with negative pressure, given by:

$$P_{DE} \approx -\rho c^2 \qquad \text{... (4.1)}$$

Quantum vacuum energy exerts a negative pressure, contributing a cosmological constant term to gravity. But both ordinary matter (atoms, molecules, and photons) and dark matter exert positive pressure, for matter:

$$P_M \approx +\rho v^2 \approx nmv^2 \qquad \text{... (4.2)}$$

v is the velocity of particles and *n* is the number density. And for radiation (photons):

$$P_\gamma = \frac{1}{3}\rho c^2 \qquad \ldots (4.3)$$

## 4.1 Change in Equation of State

The change of behaviour of missing energy density (from dark matter to dark energy) may be determined by the change in the equation of state (EOS) of a background fluid instead of a form of potential. This avoids the fine-tuning problems which is inherent in shallow potentials, almost massless fields, radiative corrections, etc. (Bento, Bertolami and Sen, 2002).

### 4.1.1 Chaplygin Gas

The modification to the assumption of no interactions and particles with zero volume which forms the basis of perfect gas is given by the van der Waals equation (Van Der Waals, 1910):

$$P = \frac{RT}{(V-b)} - \frac{a}{V^2} \qquad \ldots (4.4)$$

The constant *'b'* and *'a'* introduces the finite volume for the gas particles and strength of the interactions between them respectively. The general form is given by:

$$P = \frac{\alpha \rho c^2}{1-\beta\rho} - \gamma\rho^2 \qquad \ldots (4.5)$$

$\beta = \frac{1}{3}V_{crit}$; $\gamma = 3P_{crit}V_{crit}^2$, where, $V_{crit}$, $P_{crit}$ are the critical volume and pressure respectively. When $\beta, \gamma = 0$, the equation reduces to the usual expression:

$$P = \alpha \rho c^2 \qquad \ldots (4.6)$$

It is possible that a van der Waals type of gas could also play the role of DE (Kremer, 2007), were the interaction between DM particles can be described through a van der Waals type of gas. One can introduces a Chaplygin gas (Chaplygin, 1904), within the framework of FRW, with EOS given by, $P = -A/\rho^\alpha$, $\alpha = 1$, where, *A* is a positive constant.

The density evolves with the scale factor as:

$$\rho = \sqrt{A + \frac{B}{R^b}} \qquad \ldots (4.7)$$

This gives a direct interpolation between dust dominated phase $\rho = \sqrt{B}r^{-3}$ and de Sitter (DE) phase $P = -\rho$ through an intermediate regime described by $P = \rho$. Effective equation of state for intermediate regime is given by:

$$\rho = A^{1/(1+\alpha)} + \frac{1}{(1+\alpha)} \frac{B}{A^{\alpha/(1+\alpha)}} R^{-3(1+\alpha)} \qquad \text{... (4.8)}$$

In the equation of state, for negative pressure, we have:

$$\rho + 3P \leq 0 \qquad \text{... (4.9)}$$

If $\gamma \propto \dfrac{1}{\rho^3}$, the equation of state corresponding to that of the Chaplygin gas (Paul et al., 2010):

$$a = \frac{b}{1+a} \frac{1}{\rho_0^{1/2}} \qquad \text{... (4.10)}$$

When $\beta, \gamma = 0$ and $\alpha = -1$, the equation corresponds to that of the cosmological constant:

$$P = -\rho c^2 \qquad \text{... (4.11)}$$

For the generalised Chaplygin gas we have:

$$\rho = \rho_0 \left[ a' + \frac{(1-a')}{R^{3(1+b)}} \right]^{1/(1+b)} \qquad \text{... (4.12)}$$

$$a' = \frac{a}{\rho_0^{(1+b)}}; \quad P = -\frac{A}{\rho^{\alpha}}$$

The solution for the Chaplygin gas in the EOS gives:

$$\rho(t) = A + \rho_0 R^{-3} \qquad \text{... (4.13)}$$

The first term corresponds to dark energy and the second term to dark matter. As the universe expands (increasing R) the first term dominates. It would be asymptotically free, i.e., interaction increases with distance, if the interaction between the dark matter particles is non-abelian, which then becomes similar to a $\Lambda$ term, with negative pressure (Sivaram, Arun and Nagaraja, 2011b).

### 4.1.2 Dieterici Gas

The Dieterici equation of state is given by (Callen, 1985):

$$P = \frac{R_g T}{V-b} \exp\left(-\frac{a}{R_g T V}\right) \qquad \text{... (4.14)}$$

And the reduced coordinates are: $V_C = 2nb; \quad T_C = \dfrac{a}{4R_g b}$

And in terms of the density, we have:

$$P = R_g T \rho \exp\left(-\frac{a}{R_g T}\rho\right) \quad \ldots (4.15)$$

Using this in the continuity equation, we have:

$$\dot{\rho} + (\rho + A\rho \exp(-B\rho))\frac{\dot{R}}{R} = 0 \quad \ldots (4.16)$$

Where, $A = R_g T$; $B = \dfrac{a}{R_g T}$

On integration we get the solution as:

$$R = \exp\left(\frac{\rho^2}{2}\right) + \exp\left(\frac{1}{2} A\rho^2 \exp(-B\rho)\right) \quad \ldots (4.17)$$

We get a super-exponential term which could be relevant in the early universe, for an alternative inflation. As $T \to 0$, the second term becomes significant, i.e.:

$$P = -\frac{a}{V^2} \quad \ldots (4.18)$$

The interacting term $\propto -a'\rho^2$

Again if the interaction between DM particles is non-abelian this becomes similar to a $\Lambda$ term, with negative pressure (Sivaram and Sinha, 1979; Perkins, 2009).

For heavier dark matter particles a weak interaction coupling amongst the DM particles of $\sim 10^{-42}$ (weaker than gravity) would give the right order of DE (Sivaram, 2000; Sivaram, Arun and Nagaraja, 2011b).

We can have an acceleration equation of the expansion given by:

$$\frac{\ddot{R}}{R} = -\frac{4\pi G m^4 c^3}{3h^3}\left(1 - \frac{m_{DM}^2}{m_W^2}\right) \quad \ldots (4.19)$$

where the Planck mass associated with the weak interaction between the dark matter particles is given by: $m_W = \left(\dfrac{hc}{G_W}\right)^{1/2}$, where $G_W$ is the Fermi constant (Sivaram, Arun and Reddy, 2008; Sivaram and Arun, 2011c).

Therefore in this case: $\ddot{R}$ is positive. Therefore the expansion of the universe was repulsive hence giving the dark energy scenario (Sivaram, 1988; 2005; Sivaram and de Sabbata, 1990).

### 4.1.3 Long Range DM particle Interactions

Another approach is to consider long range interaction between dark matter particles to be described by a Lennard-Jones type of potential having both repulsive and attractive interaction (Rowlinson and Widom, 1982):

$$V \approx I\left[\frac{A}{r^{12}} - \frac{B}{r^{6}}\right] \qquad \ldots (4.20)$$

In the case of the dark matter particles the surface tension turns out to be $10^{20}$ ergs/cm$^2$, which corresponds to the dark energy density.

$$\frac{GM^2}{8\pi R^4} \approx \frac{\Lambda c^4}{8\pi G} \qquad \ldots (4.21)$$

$$\frac{Mc^2}{R^2} \approx \frac{\sqrt{\Lambda}c^4}{G} \approx 10^{20}\, ergs/cm^2 \qquad \ldots (4.22)$$

This would explain various scaling relations of large scale structures, i.e., for structures to form, the gravitational binding energy is of the order of the dark energy density, i.e. background repulsion matches inward gravitational attraction (Sivaram, 2009; Sivaram, Arun and Nagaraja, 2011b).

### 4.1.4 Viscous Dark Matter

Velten et al. (2013) have assumed CDM to possess a small bulk-viscous pressure that attenuates the growth of inhomogeneities. Based on Eckart's theory of dissipative processes, it turns out that for viscous CDM the usual Newtonian approximation for perturbation scales smaller than the Hubble scale is no longer valid, instead the neo-Newtonian approach which consistently incorporates pressure effects into the fluid dynamics correctly reproduces the general relativistic dynamics.

However, the range of validity of this result is limited only to very small values of the dark matter equation of state parameter, which corresponds to the situation where structures form. Apart from this case, Newtonian perturbation theory cannot be used to provide the correct growth of viscous CDM structures.

## 5. Alternate Models to Dark Matter and Dark Energy

As mentioned earlier, if future experiments still do not give any clue about the existence of DM, one may have to consider looking forward for alternate theories to DM. These alternatives range from modification of Newtonian dynamics and modification of Newtonian gravity to modifying the Einstein-Hilbert action. These models are still not complete and even in the modified scenarios; some amount of DM is still required to account for certain observations.

## 5.1 Modification of Newtonian Dynamics and Modification of Newtonian Gravity

The modification of Newtonian dynamics (MOND) was initially proposed as an alternative to account for the flat rotation curves of spiral galaxies, without invoking DM in the halo (Milgrom, 1983a; 1983b). The theory required an ad hoc introduction of a fundamental acceleration $\sim 10^{-10} m/s^2$.

This model proposes that the expression for force $F \; ma$ is not valid for small acceleration $(a < a_0)$ (where $a_0 \sim 10^{-10} ms^{-2}$ is the fundamental acceleration) which is seen at the outskirts of large galaxies. This gives:

$$a \quad \frac{(GMa_0)^{1/2}}{r} \qquad \ldots (5.1)$$

This gives:

$$v \quad (GMa_0)^{1/4} \qquad \ldots (5.2)$$

This being a constant gives a flat rotation curve.

Another possible explanation for the flat rotation curve of the spiral galaxies is to consider modifications to the Newtonian gravity (MONG) (Sivaram, 1994a). The Poisson's equation is given by:

$$\nabla^2 \phi \quad 4\pi G \rho \qquad \ldots (5.3)$$

where $\phi$ is the gravitational potential.

Adding gravitational self energy term to Poisson equation gives:

$$\nabla^2 \phi \quad 4\pi G \rho + k(\nabla \phi)^2 \qquad \ldots (5.4)$$

For low $\rho$, i.e. at the galaxy outskirts $\nabla^2 \phi \quad k(\nabla \phi)^2$, the solution gives the gravitational potential as: $\phi \quad k' \ln r$ and the force is given by:

$$F \quad \frac{k''}{r} \qquad \ldots (5.5)$$

Thus we get for a galaxy, $\dfrac{v^2}{r} \approx \dfrac{k''}{r}$. This implies that the velocity, v is a constant, i.e. a flat rotation curve. For Dark Energy, say constant with epoch, and cosmological constant $\Lambda$, we have:

$$\nabla^2 \phi + k(\nabla \phi)^2 \approx \Lambda c^2 \qquad \ldots (5.6)$$

In case we have just the cosmological constant (lambda) the Newtonian limit is just $\nabla^2 \phi - \Lambda c^2 \approx 0$. However it is interesting that Einstein (1917) in his original introduction of this constant got the wrong Newtonian limit of $\nabla^2 \phi + \Lambda \phi \approx 0$, which gives a Yukawa type potential falling off exponentially (Yukawa, 1935).

In principle we could have both DM and DE by having:

$$\nabla^2 \phi + k'(\nabla \phi)^2 + \Lambda c^2 \approx 0 \qquad \ldots (5.7)$$

The solution of this gives:

$$\phi \sim \ln r + k' r^2 \qquad \ldots (5.8)$$

For $r \gg r_{cluster}$, DE dominates and for $r < r_{cluster}$, DM dominates.

Modified potentials of MONG, may encounter some problems with Tully-Fisher law, which is an empirical relation between the intrinsic luminosity of a spiral galaxy and the rotation velocity, $L \propto v^4$, whereas, MOND is consistent with this law (McGaugh, 2012).

MOND claims to fit flat rotation curves of galaxies better than DM models but it fails to fit the observations of clusters of galaxies. Hot gas of clusters needs DM even with MOND and in fact observation on bullet clusters disfavours MOND (Markevitch et al., 2004).

Also the MOND theory is an ad hoc since it introduces its only parameter, the critical acceleration $a_0 \approx 10^{-10} m/s^2$, which is the introduced in an ad-hoc manner. Perhaps critical acceleration is given as $a_0 \approx cH_0$ and it is related to cosmological constant by $a_0 \approx c^2\sqrt{\Lambda}$ (Sivaram, 1994a, Milgrom, 1983a).

MOND is a modification of Newtonian mechanics rather than Newtonian gravity. The modification is written as a function $\mu$ of $\mu\left(\dfrac{g}{a_0}\right)$, where $a_0 \approx 10^{-10} m/s^2$ is a minimal acceleration obtained empirically. Thus the Newtonian acceleration $g_N$ is written as:

$$\bar{g}_N = \mu\left(\frac{g}{a_0}\right)\bar{g} = \mu(x)g \qquad \text{... (5.9)}$$

For $g \gg a_0$, we have the usual Newtonian law, where as for $\mu \ll 1$ or $g \ll a_0$, we have:

$$g = (a_0 g_N)^{1/2} \qquad \text{... (5.10)}$$

A general form of $\mu(x)$ has been given by Bekenstein (2004) as

$$\mu(x) = \frac{\sqrt{1+4x}-1}{\sqrt{1+4x}+1} \qquad \text{... (5.11)}$$

so that the generalised acceleration due to gravity is now:

$$g = -\frac{GM}{R^2} - \frac{(Ga_0)^{1/2} \mu^{1/2}}{R} \qquad \text{... (5.12)}$$

where, $M$ and $R$ are the mass and radius of any gravitating object. Thus:

$$g = g_N + (a_0 g_N)^{1/2} \qquad \text{... (5.13)}$$

This can also be written as:

$$g = g_N (1+\eta) \qquad \text{... (5.14)}$$

where $\eta = \left(a_0/G\right)^{1/2} \left(R/M^{1/2}\right)$, $a_0 \approx 10^{-10} \, cm/s^2$ (by coincidence $a_0 \approx cH_0$, in Sivaram (1994a) it was noted that $a \approx c\sqrt{\Lambda}$, where, $\Lambda$ is the cosmological constant of dark energy).

When $\left(R/M^{1/2}\right)$ is small, we have the Newtonian value, and MOND is valid if $\left(R/M^{1/2}\right)$ is large.

Now for globular clusters, $M \sim 10^6 M_{sun}$, $R \sim 20\,pc$, $\eta \sim 0.01$, so Newtonian law is almost valid. For giant spirals like Milky Way, $M \sim 10^{11} M_{sun}$, $R \sim 10\,kpc$, $\eta \sim 2$. So MOND explains flat rotation curve without dark matter. Thus the (constant) flat rotational velocity (independent of $R$) is:

$$v = (GMa_0)^{1/4}; \quad M \propto v^4 \qquad \text{... (5.15)}$$

MOND has supposedly proved more successful than DM model, in explaining flat rotation curves of galaxies, but does not work for large galaxy clusters, where some DM is required anyway.

The recently postulated Neptune-sized planet with an orbital period of 15,000 years around the sun could provide us with a testing ground for modification of Newtonian dynamics (Sivaram, Arun and Kiren, 2016). For the solar system the gravitational acceleration (dominated by the sun's mass) drops to $a_0$, at a distance from the sun of $r \approx 10^{15} m$ (a few thousand AU). This is at about the distance Planet Nine is expected to orbit, where the gravitational interaction of Jupiter, Saturn, etc. and also the effects of the nearest stars would be far less.

So if there are objects orbiting the sun, at a few thousand AU (or beyond), MOND would suggest that their orbital velocity tends to a constant value given as,

$$v_C \approx (GM_{sun}a_0)^{1/4} \approx 0.5 km/s \qquad \ldots (5.16)$$

independent of their distance from the sun. This is testable by techniques such as space Doppler.

## 5.2 Tensor–Vector–Scalar Theory

The tensor-vector-scalar (TeVeS) theory was developed by Bekenstein (2004; 2005), as a relativistic generalisation of MOND, and is derived from the action principle. In the weak-field approximation, static solution, TeVeS gives the MOND acceleration formula. This theory can account for the gravitational lensing, since it is a relativistic theory.

MOND is not a complete theory (it violates the law of conservation of momentum), but, such conservation laws are satisfied if they are derived using the action principle. The Lagrangian for the non-relativistic limit is given by:

$$L = -\frac{a_0^2}{8\pi G} f\left(\frac{|\nabla \Phi|^2}{a_0^2}\right) - \rho \Phi \qquad \ldots (5.17)$$

where $\rho$ is the mass density, $a_0$ is a scale of acceleration.

A relativistic MOND theory seems essential if gravitational lensing by extragalactic systems and cosmology are to be understood without reliance on DM. The Lagrangian contains, in addition to the Einstein–Hilbert action for the metric field $g_{\mu\nu}$, terms pertaining to a unit vector field $u^\alpha$ and two scalar fields. The action is therefore given by:

$$S_{\text{TeVeS}} = \int (L_g + L_S + L_V) d^4 x \qquad \ldots (5.18)$$

Where, $L_g = -\frac{1}{16\pi G} R\sqrt{-g}$

Even though this theory claims to explain phenomena, such as gravitational lensing, the theory is unable to account for phenomena for galactic dynamics and lensing. Also the observation of Bullet Cluster (of a pair of colliding galaxy clusters) is not compatible with any current modified gravity theories (Angus, Famaey and Zhao, 2006; Clowe et al., 2006).

TeVeS theory has been shown to explain galaxy curves, galaxy distributions and CMBR data (Skordis et al., 2006). The central baryonic surface density appears correlated with the core radius, i.e. there is close correlation between enclosed surface densities of luminous and dark matter in galaxies (Gentile et al., 2009). As these results suggest that a single scale governs dynamics of all galaxies it is difficult to reconcile this with the picture of DM as particles.

MOND type theories however have already a low acceleration scale $a_0$ leading to a close agreement with observed luminous surface densities (Donato at al., 2009). Thus although search for DM particles (of unknown type) continues diligently, it is not clear or certain that it is the unequivocal answer to the DM problem.

## 5.3 Modification of Einstein-Hilbert Action

Another possible way in which the need for DM may be avoided can be seen from modifying the Einstein-Hilbert action. In GR, the Einstein-Hilbert action, when varied to obtain equations of motion for the space-time metric, yields the Einstein's field equations. The action $S[g]$ is given by the integral of the Lagrangian. $S[g]$ gives rise to the vacuum Einstein equations:

$$S[g] = \int \kappa R \sqrt{-g} \, d^4x \qquad \ldots (5.19)$$

Where, $g = |g_{ab}|$ is the determinant of space-time Lorentz metric, $R$ is the Ricci scalar, $\kappa = c^2/16\pi G$ is constant, the Lagrangian is $R\sqrt{-g}$, and the integration is done over a region of space-time.

The Lagrangian need not necessarily be of the form $R\sqrt{-g}$. Instead the Lagrangian could be a function of the Ricci scalar, with the form $f(R)\sqrt{-g}$. This yields a field equation given by:

$$f'(R)R_{\mu\nu} - \frac{1}{2}g_{\mu\nu}f(R) = \frac{8\pi G}{c^4}T_{\mu\nu} \qquad \ldots (5.20)$$

where $f(R)$ can have the form given by:

$$f(R) = \frac{R}{\left(1 - R/R_{max}\right)\left(1 - L_{Pl}^2 R\right)} \qquad \ldots (5.21)$$

where $L_{Pl}^2 = \frac{1}{R_{max}}$ gives the maximal curvature. The field equation is then obtained as:

$$R_{\mu\nu} - \frac{1}{2} g_{\mu\nu} R = \frac{8\pi G}{c^4} T_{\mu\nu} \qquad \ldots (5.22)$$

The motivation for choosing the form as given above is by analogy with the Born-Infeld modification of Maxwell's electrodynamics. Born and Infeld, tried to avoid the self energy divergence of a point charge in classical electrodynamics by incorporating (introducing) a maximal field strength $E_{max}$ and modifying the classical electrodynamic action as:

$$\alpha_E = \frac{E_{max}^2}{4\pi}\left[1 - \sqrt{1 + \left(\frac{F_{\mu\nu} F^{\mu\nu}}{2E_{max}}\right)^2}\right] \qquad \ldots (5.23)$$

In the limit of fields $E \ll E_{max}$, it reduces to the usual $\frac{1}{16\pi} F_{\mu\nu} F^{\mu\nu}$.

The self energy calculated with the modified theory gives a finite value for the self energy and a minimal length scale as $r_{min}^2 = e/E_{max}$, $e$ being the electron charge. Here the curvature is the equivalent to the field strength, and so the maximal curvature $R_{max} = 1/L_{Pl}^2$ plays the role of $E_{max}$ in the Born-Infeld theory. Using this we get the field equation as:

$$\frac{R_{\mu\nu}}{\left(1 - L_{Pl}^2 R\right)^2} - \frac{2R_{\mu\nu}}{\left(1 - L_{Pl}^2 R\right)} = \frac{8\pi G}{c^4} T_{\mu\nu} \qquad \ldots (5.24)$$

On rearranging the terms we get:

$$R_{\mu\nu} - 2R_{\mu\nu}\left(1 - L_{Pl}^2 R\right) = \frac{8\pi G}{c^4} T_{\mu\nu}\left(1 - L_{Pl}^2 R\right)^2 \qquad \ldots (5.25)$$

As $R \to R_{max}$; $R_{max} \to 1/L_{Pl}^2$. Then the above equation becomes $R_{\mu\gamma} = 0$. This shows the non existence of singular states.

In the context of accounting for DE we consider a cosmic background with a minimal curvature by accordingly modifying the Einstein action. Analogous to the above case, for minimum curvature $R_{min}$, GR can be modified as:

$$f(R) \approx \frac{R}{\left(1 - R_{\min}/R\right)} \qquad \ldots (5.26)$$

If $R \gg R_{\min}$, it reduces to the usual GR.

In the expansion, i.e. $f(R) \approx R + a' R_{\min} + b\frac{R_{\min}^2}{R} + \ldots$, the first term represents GR, the second term is a background minimal curvature, equivalent to $\Lambda$, which shows the presence of DE. The higher order terms involving $R_{\min}^2$ are too small to significantly contribute to the local behaviour, so with the first two terms it may be difficult to distinguish this theory from GR with a cosmological constant.

There are $1/R$ theories that have been suggested, which do not require DE but have accelerating universe. These alternative gravity theories that do not require some unknown DE or DM should satisfy all tests of GR as well as match with observations. There are problems with galaxy formation in MOND (Sivaram 1994a). These theories also have severe constraints from CMB anisotropy and lensing. So far there are no complete theories, although string theory could give scalar tensor theories in low energy limit.

In general, these models are equivalent to GR plus massive scalar fields. Solar System tests for relativistic theories of gravity include gravitational redshift, deflection of light by the Sun, and planetary orbit precession at perihelion and GR is consistent with these experimental tests. Any correction to the Newtonian potential has to satisfy the constraints on equivalence principle and solar system observations.

Also, the new gravitationally-induced interactions lead to observable effects at microscopic and macroscopic scales. These facts make very unlikely the viability of $f(R)$ models in accounting for the change in the late cosmic evolution (Olmo, 2007), apart from difficulties with equivalence principle, as $f(R)$ theories are necessarily equivalent to scalar-tensor theories (Sivaram and Campanelli, 1992a; 1992b).

Gravitational wave astronomy, starting with the event GW150914 (LIGO Scientific Collaboration and Virgo Collaboration, Abbott et al., 2016), could be fundamental for discriminating various theories of gravity. With improved sensitivity and advanced detection of gravitational waves, then the accurate angle- and frequency-dependent response functions of interferometers for gravitational waves arising from various theories of gravity, i.e. general relativity and extended theories of gravity will be the definitive test for general relativity, and help in discriminating among various gravity theories (Corda, 2009).

### 5.4 Can MOND be differentiated from Newtonian DM Theory?

As seen above, two poplar theoretical concepts to resolve the apparent inconsistency of Newtonian dynamics over galactic scales and beyond is assumption of ubiquitous presence of DM and secondly the assumption that Newtonian gravitational law or dynamics requires modification.

Dunkel (2004) argues that for a system satisfying a fixed relationship between gravitational fields caused by DM and visible matter, a generalised MOND equation reducing to the usual MOND law can be formulated. Thus MOND is interpreted as a special case of DM theory.

Briefly both gravitational potentials $\phi_v(x), \phi_d(x)$ due to visible and DM satisfying Poisson equations, $\nabla^2 \phi(v,d) \; 4\pi G \rho(v,d)$, for both visible and DM leading to the equations of motion:

$$m\ddot{x} \; -m\nabla[\phi_v(x)+\phi_d(x)] \qquad \text{... (5.27)}$$

Defining the accelerations $g(v,d)(x) \; -\nabla \phi(v,d)x$, this is written as, $\ddot{x} \; g(v)+g(d) \; g$, and if further assumed that these acceleration vectors point in the same direction:

$$\ddot{x} \; \left(1+\frac{g(d)}{g(v)}\right)g_v \qquad \text{... (5.28)}$$

With $g_v \; g - g_d \geq 0$, we have:

$$\ddot{x} \; \left(1+\frac{1}{g/g_d - 1}\right)g_v \qquad \text{... (5.29)}$$

$$g_v \; \left(\frac{\varepsilon}{\varepsilon+1}\right)g \; \mu(\varepsilon)g, \; \varepsilon(x) \; \frac{g(x)}{g_d(x)}-1 \geq 0 \qquad \text{... (5.30)}$$

So we have a generalisation of the MOND equation,

$$g_v \; \mu\left(\frac{g}{a_0}\right)g \qquad \text{... (5.31)}$$

Here, instead of the constant $a_0$ of MOND, we have the acceleration field $a(x)$ defined as,

$$\varepsilon(x) \equiv \frac{g(x)}{a(x)} \equiv \frac{g(x)}{g_d(x)} - 1 \qquad ...(5.32)$$

$$\frac{1}{a(x)} \equiv \frac{1}{g_d(x)} - \frac{1}{g(x)}, \quad g(x) \equiv g_v(x) + g_d(x) \qquad ...(5.33)$$

MOND is the special case where $a(x) \equiv a_0$, which implies a fixed relation between acceleration fields due to visible and DM. It can be seen from the above relation that for $a_0$ to be small, galaxies satisfying MOND limit are DM dominated. (Again $\varepsilon(r) \to 0$, as $g(r) \to g_d(r)$, i.e. DM dominated, the Tully-Fisher law also follows.)

Current DM models cannot explain in which cases $g_v \ll g_d$ and $a_\infty > 0$ $\left(a_\infty \equiv \lim_{r \to \infty} a(r)\right)$, are satisfied and so modifications even of conventional DM theory are required.

Milgrom's MOND is equivalent to having an additional potential $\phi_d$ (generated by DM) to the usual Newtonian dynamics. Again in the case of galaxy clusters, the X-ray emissions require substantial hot gas which has to be gravitationally supported by DM and cannot be satisfactorily explained by MOND. So presence of at least some DM is required.

## 5.5 Dark Energy and Modified Gravity

Strong observational evidence that the universe is accelerating has again led to the standard explanation involving an unknown dark energy component. Theoretical problems with such scenarios (like fine tuning, presence of a $\Lambda$ term with just such a value at present epoch etc.) have also led to suggestions that GR be modified in such a way that leads to observed accelerated expansion. Can the two alternative scenarios be observationally distinguished?

Kunz and Sapone (2007) have shown that a generalised DE model can for instance match the growth rate in the DGP (Dvali-Gabadadze-Porrati) model and reproduce the 3+1 dimensional metric perturbations. Cosmological observations cannot distinguish the two cases.

In DGP model (Dvali, Gabadadze and Porrati, 2000), a popular model as an alternative to DE, in which gravity leaks off a 4D Minkowski-brane into 5D space-time. On small scales gravity is bound to the 4D brane giving Newtonian gravity. The Hubble expansion evolves as:

$$H^2 - \frac{H}{R_C} = \frac{8\pi G}{3}\rho_m \qquad \text{... (5.34)}$$

$R_C$ crossover scale separates 5D and 4D regimes. This scale $\sim 1/H_0$, and generates late time acceleration.

Comparing this to usual Friedmann equation with additional DE component, the crossover term can be shifted to the RHS and be thought of as a DE contribution:

$$\rho_{DE} = \frac{3H}{8\pi G R_C} \qquad \text{... (5.35)}$$

This gives rise to the effective DE equation of state:

$$1 + w_{DE} = -\frac{\dot{H}}{3H^2} \qquad \text{... (5.36)}$$

Thus impossible to rule out DE based on measurements of expansion history of the universe like in the SN-Ia data. Present observations show that even growth rate of structures cannot differentiate between the alternatives. Again construction of a matching DE model for DGP case is very fine tuned.

Ironically as current observations seem to indicate $w_{DE} \approx -1$ (a $\Lambda$ term) it is rather the modified GR models (with all their fine tuning problems) that are likely to be ruled out.

## 6. Summary and Outlook

In this review, we have given an overview of the current understanding of Dark Matter and Dark Energy, along with the possible alternative theories. The only observational evidence we have so far is that we require some amount of DM to account for certain observations, but we do not yet understand the nature of these particles. The proposed candidates range from WIMPS and Axions to exotic particles. This review covers the entire spectrum of these DM candidates highlighting the different detection techniques.

The question of DE is also far from resolved. The possible models include cosmological constant, quintessence, phantom energy, etc. The current observations is consistent with a cosmological constant playing the role of repulsive gravity. But the models do not predict why DE constitutes close to 70% of the universe's energy density. And present observations show that even growth rate of structures cannot differentiate between the alternatives.

We also consider another class of theories that attempts to unify DM and DE as a single phenomenon where the change of behaviour of missing energy density, from DM to DE may be determined by the change in EOS of a background fluid instead of a form of potential.

In the absence of any positive detection of DM in recent experiments, we may also have to start looking at the alternatives seriously. Here we included the various alternatives that are in vogue currently and look at the advantages and drawbacks compared to the DM models. There are many detectors and ingenious experiments ongoing to detect these elusive particles.

The next few years are sure to see a lot of excitement in these areas. It looks to be a very promising and challenging times ahead; is it going to be dark matter or one of the alternatives, or something entirely new.